\documentclass[11pt]{article}

\usepackage{authblk}
\usepackage{hyperref}
\usepackage{amsfonts}
\usepackage{amsmath}
\usepackage{amssymb}
\usepackage{siunitx}
\usepackage{booktabs}
\hypersetup{
    colorlinks=true,
    linkcolor=blue,
    filecolor=magenta,      
    urlcolor=cyan,
}
\usepackage{sectsty}
\usepackage{graphicx}
\usepackage[
backend=biber,
style=numeric,
sorting=none
]{biblatex}
\usepackage{float}

\title{\textbf{FRET-guided selection of RNA 3D structures}}

\author[1]{Mirko Weber}
\author[1]{Felix Erichson}
\author[2,3]{Maciej Antczak}
\author[3]{Tomasz Zok}
\author[4]{Fabio D. Steffen}
\author[2,3]{Marta Szachniuk}
\author[1,*]{Richard Börner}

\affil[1]{Laserinstitut Hochschule Mittweida, University of Applied Sciences Mittweida, Technikumplatz 17, 09648, Mittweida, Germany}
\affil[2]{Institute of Bioorganic Chemistry, Polish Academy of Sciences, Noskowskiego 12/14, 61-704, Poznan, Poland}
\affil[3]{Institute of Computing Science, Poznan University of Technology, Piotrowo 2, 60-965, Poznan, Poland}
\affil[4]{Department of Oncology, University of Zurich, University Children's Hospital, 8008, Zurich, Switzerland}
\affil[*]{Corresponding author: \texttt{richard.boerner@hs-mittweida.de} - Tel: +49 3727 58 1009}

\begin{document}
\maketitle	


\noindent\textbf{Abstract}\\
Integrative biomolecular modeling of RNA relies on structural refined collections and accurate experimental data that reflect binding and folding behavior. However, the prediction of such collections remains challenging due to the rugged energy landscape and extensive conformational heterogeneity of large RNAs. To overcome these limitations, we applied a FRET-guided strategy to identify RNA conformational states consistent with single-molecule FRET (smFRET) experiments. We predicted 3D structures of a ribosomal RNA tertiary contact comprising a GAAA tetraloop and a kissing loop using three popular RNA 3D modeling tools namely RNAComposer, FARFAR2, and AlphaFold3, yielding a collection of candidate conformations. These models were structurally validated based on Watson-Crick base-pairing patterns and filtered using an eRMSD threshold. For each retained structure, we computed the accessible contact volume (ACV) of the sCy3/sCy5 dye pair using FRETraj to predict FRET distributions. These distributions were then compared and weighted against experimental smFRET data to identify conformational states compatible with the observed FRET states. Our results demonstrate that experimental transfer efficiencies can be reproduced using \textit{in silico} predicted RNA 3D structures. This FRET-guided workflow, combined with structural validation, lays the foundation for capturing the highly diverse conformational states characteristic of flexible RNA motifs.

\newpage

\section{Introduction}
Three-dimensional (3D) structure prediction and molecular dynamics (MD) simulations have been instrumental in resolving biomolecular structures at atomic resolution \cite{Karplus2002}. While these methods have been particularly impactful in protein structure prediction and dynamics \cite{Abramson2024, Nam2023, Hospital2015}, attention is increasingly shifting towards RNA and RNA-protein complexes, due to their structural complexity and functional diversity \cite{Statello2021, Schug2019}. In response, RNA 3D structure prediction has become a rapidly evolving field, with sustained efforts aimed at enhancing accuracy, reliability, and robustness. A key development in this area was the launch of RNA-Puzzles \cite{Cruz2012}, a community-wide initiative that facilitates blind benchmarking of RNA 3D prediction tools against experimentally determined reference structures \cite{Miao2015, Miao2017, Miao2020, Bu2024}. This effort has significantly accelerated the field by enabling objective performance evaluation and encouraging methodological advancements \cite{Bernard2024}. Despite a growing range of computational strategies, modeling RNA structures remain dependent on experimental data from high-resolution techniques such as X-ray crystallography \cite{Galli2014}, NMR spectroscopy \cite{Ferguson1967}, or cryo-EM \cite{Subramaniam2022}. As the demand for accurate modeling of flexible and functionally relevant RNA conformations continues to grow, the integration of predictive modeling with experimental validation, such as FRET, remains a cornerstone of progress in RNA structural biology.

Although 3D structure prediction and MD simulations provide critical insights into the three-dimensional organization of RNA, they often focus on identifying a single, energetically favorable state or sample a limited subspace of the full conformational landscape. This single-state representation is inadequate to capture the structural heterogeneity observed in many functional RNAs, where multiple transiently populated conformations co-exist. \cite{Sponer2018}. To address these limitations, in-solution techniques such as SAXS \cite{Svergun2003, Tants2023}, EPR \cite{Ponce-Salvatierra2019, Jeschke2018}, and FRET \cite{Forster1948, Lerner2021} serve as complementary approaches for probing conformationally heterogeneous RNA structure collections. These methods enable structural characterization and are particularly well-suited for detecting dynamic behavior \cite{Hellenkamp2018, Peter2022}. EPR and single-molecule FRET (smFRET), in particular, provide access to highly precise, site-specific distance information that is critical for resolving low-populated or transient states. Within integrative or hybrid modeling approaches \cite{Dimura2016, Vallat2018}, such distance constraints offer an experimental benchmark for selecting and validating conformations from predicted structure collections. Among available experimental techniques, smFRET combines nanometer-scale resolution with single-molecule sensitivity, enabling direct comparison between predicted and experimental distances or whole distance distributions \cite{Steffen2024}, which is an ideal technique for detecting conformational subpopulations and conformational changes to capture the dynamic behavior of RNAs.

RNA folding is a dynamic, hierarchical process, where proteins assist folding by stabilizing native-like intermediates, and tertiary interactions are essential for compact RNA architecture \cite{Woodson2011}. In particular, tertiary contacts between distant secondary structure elements frequently form only transiently or in the presence of stabilizing factors such as Mg (II) ions or RNA-binding proteins \cite{Leontis2001,Woodson2010,Gerhardy2021}. This structural plasticity, i.e., the intrinsic flexibility of RNA, defines a collection of unbound states. Characterizing this conformational collection experimentally remains challenging, due to its high structural heterogeneity. Single-molecule techniques such as smFRET offer a powerful approach to probe this heterogeneity \cite{Lerner2021}, as they can resolve subpopulations within the collection and guide structure prediction by identifying conformations consistent with observed distance constraints \cite{Dimura2020, Steffen2024, Hanke2024}.

\begin{figure*}[!b]%
\centering
\includegraphics[width=\textwidth]{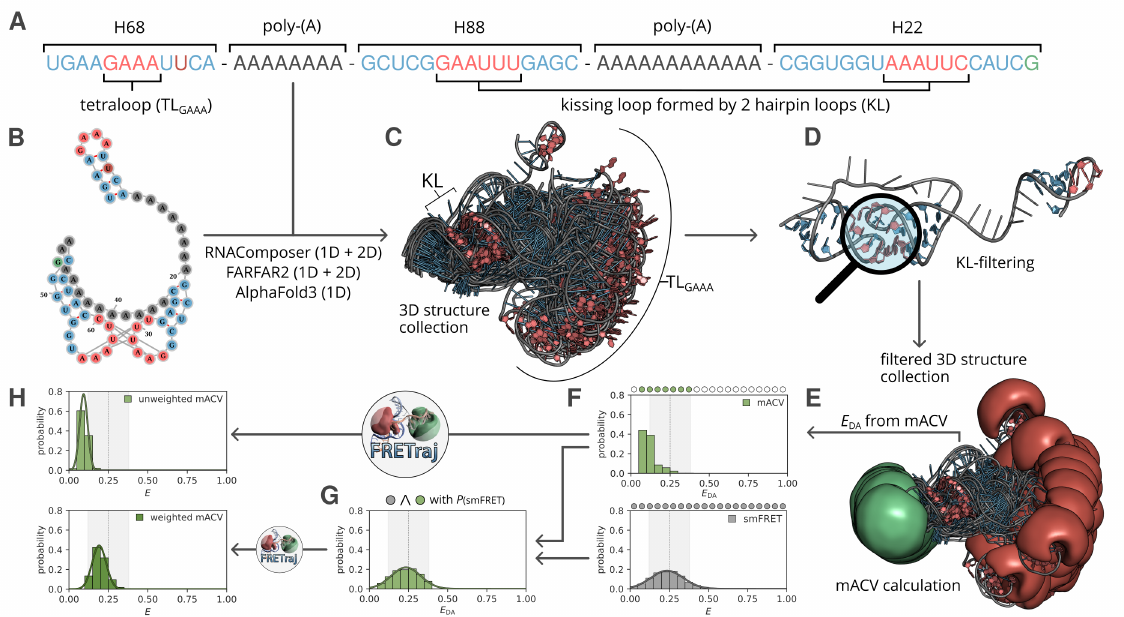}
\caption{Representation of the workflow from sequence and secondary structure to the modeling of a final 3D structure collection that best represents the experimental smFRET data under the condition of a correctly folded KL. \textbf{(A)} Sequence showing all structure elements of the investigated model construct. \textbf{(B)} Secondary structure representation of the construct. \textbf{(C)} A complete structure collection of 1,000 structures predicted by RNAComposer. \textbf{(D)} Evaluation of predicted structures with Barnaba to ensure correctly folded KLs in each structure with accurate WC base-pairing and the calculation of the eRMSD to the reference structure. \textbf{(E)} For each filtered structure, the mACVs of the dyes sCy3 and sCy5 were predicted, and the mean FRET efficiency was calculated. \textbf{(F)} Distribution of the \ensuremath{E_{\mathrm{DA}}} values calculated from the mACVs alongside the experimental smFRET distribution, each represented with their respective bin populations. \textbf{(G)} Reweighting of the structure collection, ensuring that structures are selected from each populated bin with the same probability as observed in the smFRET distribution. \textbf{(H)} The comparison of FRET efficiencies after photon sampling with FRETraj for both the unweighted \ensuremath{E_{\mathrm{DA}}} distribution and the weighted distribution with the probabilities from the experiment.}\label{fig1}
\end{figure*}

Here, we apply FRET-guided integrative modeling to an ribosomal RNA model construct, featuring a kissing loop (KL) and highly flexible GAAA tetraloop (TL\textsubscript{GAAA}) domain, which can serve as a tertiary contact binding to the KL \cite{Gerhardy2021}, that we aim to characterize structurally in its unbound state. Using RNAComposer \cite{Popenda2012, Sarzynska2023}, FARFAR2 \cite{Watkins2020}, and AlphaFold3 \cite{Abramson2024} we generated a collection of RNA 3D models based on diverse structural prediction strategies. For model validation, we calculated multiple ACV \cite{Steffen2016, Steffen2021, Steffen2024} to predict \textit{in silico} FRET from the dye-labeled RNA models and used these to identify an initial pool of candidate structures matching the experimental smFRET distribution. To ensure structural plausibility, we applied established validation metrics, including Watson–Crick (WC) base-pairing \cite{Leontis2001} analysis and the eRMSD score \cite{Bottaro2014, Bottaro2019}, yielding a refined collection suitable for smFRET-guided conformer selection.

Ultimately, we used the experimentally measured smFRET distribution to guide structure selection from the validated collections. Instead of comparing FRET values post hoc, we applied the distribution as a filter to extract conformers whose ACV-based predictions matched the experimental data. By enabling targeted sampling within the conformational landscape resolved by smFRET, this approach advances RNA structural modeling beyond static representations toward a dynamic, experimentally grounded collection.

\section{Methods}\label{sec2}
\subsection{Probing the unbound state with smFRET\label{subsec1}}
smFRET measurements of the Cy3/Cy5-labeled ribosomal RNA model construct, designed to probe KL formation and potential TL\textsubscript{GAAA} binding, were performed and analyzed according to standard protocols \cite{Steffen2024, Hellenkamp2018}. The construct was prepared as described in \cite{Gerhardy2021} (Figure \ref{fig1}A and B). For molecular sorting POE-type smFRET experiments were performed in standard buffer containing [K$^{+}$] = \SI{116}{\milli\mole\per\liter} and EDTA to allow KL formation without TL\textsubscript{GAAA} binding. The smFRET data was further corrected for background, bleed through, and direct excitation. $\gamma$-correction was applied to account for differences in detection efficiency and quantum yield of the FRET pair. These corrections are essential for the comparability of the experimental data with FRET distributions derived from \textit{in silico} structure collections using FRETraj \cite{Steffen2021}, together with the experimental burst size distribution (Supplementary Figure 3), i.e., the sum of donor and acceptor intensities (Figure \ref{fig1}F).

\subsection{\textit{In silico} RNA 3D structure prediction and MD simulation\label{subsec2}} 
Structure collections consisting of 10,000 structures each were predicted using RNAComposer, FARFAR2 and AlphaFold3 (Supplementary Methods). Additionally, we performed six independent \SI{1}{\micro\second} MD simulations with different initial seed structures chosen to represent structurally diverse conformations (Supplementary Methods). An equal number of structures was sampled from each of the six MD simulations for comparison with the 3D prediction tools and for subsequent use in the FRET-guided reweighting. ACVs of both dyes were calculated for each structure predicted by the 3D prediction tools as well as every \SI{100}{\pico\second} along the MD trajectories. Photon emission events were simulated by FRETraj \cite{Steffen2021, Hoefling2011, Hoefling2013}. The FRET distributions were also analyzed both individually for each MD simulation (Supplementary Figure 2) and as a combined collection (Figure \ref{fig4}A and B - MD simulation). All MD simulations were annotated every \SI{100}{\pico\second} based on Watson–Crick (WC) base-pairing using the Barnaba toolbox \cite{Bottaro2019}, resulting in six sets of 10,000 annotated structures representing the unbound RNA conformation (averaged in Figure \ref{fig3}A right).

\subsection{Estimating structure collection size with the Kullback-Leibler divergence\label{subsec3}} 
\begin{figure*}[!b]%
\centering
\includegraphics[width=0.95\textwidth]{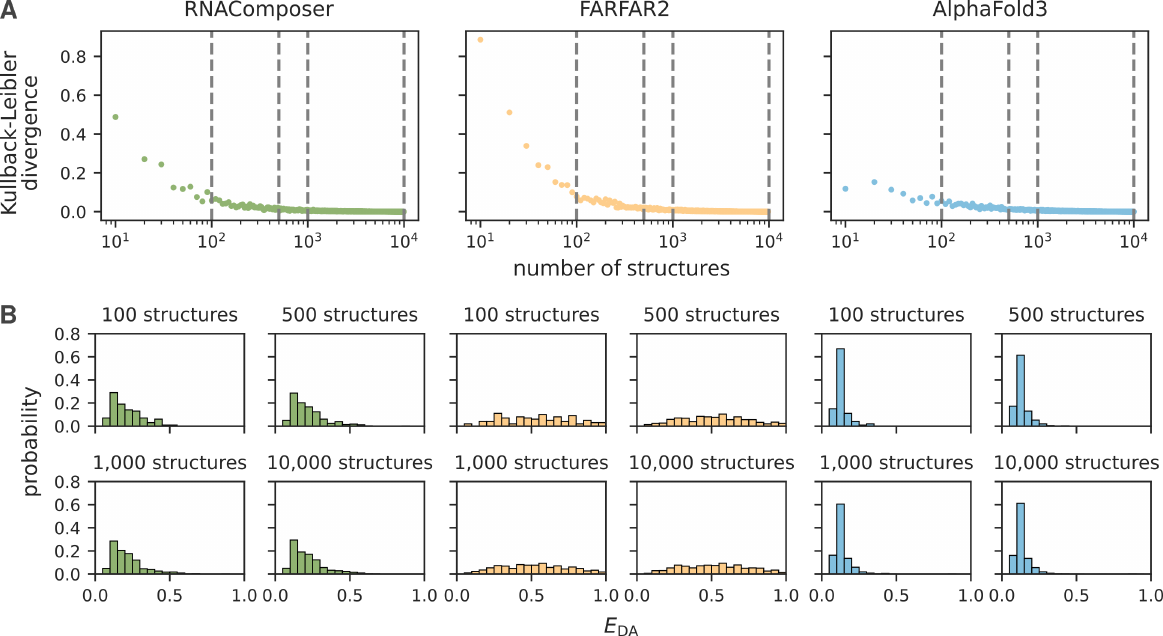}
\caption{Distribution of FRET efficiencies calculated from mean donor–acceptor distances via ACVs for RNAComposer, FARFAR2, and AlphaFold3, shown alongside the initial reference distribution (10,000 structures). \textbf{(A)} KLD between subsets of 10 structures and the full 10,000 structure collections for all three tools. Dashed vertical lines indicate the subset sizes used for the \ensuremath{E_{\mathrm{DA}}} distributions shown in B. \textbf{(B)} The distributions based on 10,000 structures denote the reference probability distributions.
}\label{fig2}
\end{figure*}

To estimate the minimum number of structures required for our structure collections to be representative of a collection of $N = 10,000$ structures, we compared the full structure collections probability distribution \( P \) of the mean transfer efficiencies \(\langle E_{\mathrm{DA}} \rangle\) calculated via ACVs with subdistributions sampled \( Q_{x_i} \) of each prediction tool. For each \( i \), where \( i \in \{10, 20, 30, \dots, 9990\} \), \( x_i \) represents the number of samples drawn from the structure collection of size \( N\) and for each \( x_i \) the structures were randomly sampled from \(N\). To compare the probability distribution we applied the Kullback-Leibler divergence (KLD) \cite{Kullback1951}, thus for each repetition it is computed as:
$$
D_{\text{KLD}}(P \| Q_{x_i}^{(r)}) = \sum_{j=1}^m p_j \log\left(\frac{p_j}{q_j^{(r)}}\right),
$$
where \( m \) is the total number of bins, \( p_j \) is the true probability of bin \( j \), and \( q_j^{(r)} \) is the corresponding probability of the bin from the sampled distribution.
The mean KLD across all repetitions for a given \( x_i \) is then defined as:
$$
\overline{D_{\text{KLD}}}(x_i) = \frac{1}{k_i} \sum_{r=1}^{k_i} D_{\text{KLD}}(P \| Q_{x_i}^{(r)}).
$$
where \( k_i\) was increased until the standard deviation of the corresponding KLD values was less than 1\% of their mean. Based on this information, we determined the minimum number of structures required for each structure collection to achieve an accurate representation (Figure \ref{fig2} and Supplementary Figure 1). 
 
\subsection{Validation of the KL formation and filtering of the structure collections}\label{subsec4}
 
To ensure formation of the KL as observed in the reference structure (PDB-ID: 3JCT and Figure \ref{fig3}A), we calculated the WC base-pairs annotated by Barnaba \cite{Bottaro2019} across all predicted structures with those from 10,000 MD-derived states. This served to validate the structural preservation of the KL motif. Subsequently, WC base-pairs were annotated for all structures from RNAComposer, FARFAR2, and AlphaFold3. Structures were classified as exhibiting preserved KL formation if all relevant base-pairs were canonical, except U27–A59, which was allowed to vary consistent with the reference structure. Any deviation led to annotation as altered KL formation (Figure \ref{fig3}C). In parallel, the structural similarity of each model to the reference was quantified using the eRMSD \cite{Bottaro2014}. Models with \(\text{eRMSD} \leq 0.8\) and preserved base-pairing were selected for further analysis (Figure \ref{fig3}B left). To assess structural stability and confirm base-pair preservation after refinement, each filtered model was subjected to a short \SI{1}{\nano\second} MD simulation, followed by re-annotation and recomputation of WC base-pairs and eRMSD (Figure \ref{fig3}B right).

\subsection{Reweighting of FRET efficiencies from mACVs}

For the comparison of the structure collections obtained from the investigated 3D prediction tools and the described MD simulations with an experimental smFRET distribution, photon emission events were simulated using FRETraj based on two different structure sampling strategies. In the first approach, each structure was selected exactly once, thereby preserving the original distribution of predicted mean FRET efficiencies \( E_{\mathrm{DA}} \). Consequently, the relative frequency of structures in each bin \( B_i \) reflected the number of structures falling within the corresponding \( E_{\mathrm{DA}} \) range, i.e.,
\[
P_{\text{unweighted}}(i) = \frac{|B_i|}{N},
\]
where \( |B_i| \) denoted the number of structures in bin \( i \), and \( N \) was the total number of structures in the filtered collection (Figure \ref{fig1}H - unweighted mACV).

In the second approach, structures were sampled such that the resulting distribution approximated the experimental smFRET distribution \( p^{\text{smFRET}} \) (Figure \ref{fig1}H - weighted mACV):
\[
P_{\text{weighted}}(i) =
\begin{cases}
p_i^{\text{smFRET}} & \text{if } |B_i| > 0, \\
0 & \text{otherwise,}
\end{cases}
\]
where \( p_i^{\text{smFRET}} \) denoted the smFRET probability of bin \( i \), provided that \( B_i \) was not empty.

Based on this, two scenarios were distinguished for each bin:
\paragraph{Scenario 1:} If the relative number of structures in \( B_i \) exceeded the experimental probability \( p_i^{\text{smFRET}} \), a subset of structures was randomly sampled from \( B_i \) to match \( p_i^{\text{smFRET}} \).

\paragraph{Scenario 2:} If the number of structures in \( B_i \) was insufficient to satisfy \( p_i^{\text{smFRET}} \), all structures in \( B_i \) were selected repeatedly until the required number was reached. Any remaining fraction was filled by randomly sampling additional structures from \( B_i \). Each repetition was randomly permuted to construct an expanded set \( \bar{B}_i \), from which sampling was performed to match \( p_i^{\text{smFRET}} \).

\section{Results}

\subsection{1,000 RNA 3D structures of the KL-TL\textsubscript{GAAA} are sufficient to capture experimental FRET distributions}

Today, a wide range of \textit{in silico} RNA 3D structure prediction tools is available, each employing different methodologies for structure modeling. However, they all share a key characteristic: the ability to generate RNA 3D structures with exceptional speed and efficiency. Nevertheless, it remains unknown how many structures are actually needed to accurately represent a diverse structure collection, particularly in the context of FRET-guided integrative modeling. 

We compared three widely used tools, RNAComposer, FARFAR2, and AlphaFold3, and evaluated their ability to predict the unbound state of the KL-TL\textsubscript{GAAA} model construct. Each tool was used to generate 10,000 structures, serving as a baseline for evaluating the structural diversity of the unbound state (Figure \ref{fig2}). We generated subsets of varying sizes from each baseline dataset and assessed their similarity to the full \ensuremath{E_{\mathrm{DA}}} distribution using the KLD.

Therein, the bin size of the FRET distribution is crucial. The more precisely we represent the FRET values the more structures are required to accurately resemble the baseline. A FRET bin size of 0.05 was chosen based on a benchmark study demonstrating an experimental accuracy of $\Delta$E=$\pm$0.05 \cite{Hellenkamp2018}, and is used as a criterion throughout this work. Our results demonstrat, that the KLD converges toward zero across all \ensuremath{E_{\mathrm{DA}}} bin sizes tested (Supplementary Figure 1). The minimum feasible subset is determined by the diversity of the original \ensuremath{E_{\mathrm{DA}}} distribution. As a result, FARFAR2 requires more structures compared to RNAComposer and AlphaFold3 to fulfill the FRET criterion and to reproduce the baseline (Figure \ref{fig2}A). RNAComposer and AlphaFold3 show a starting convergence at a subset size of around 100 structures, while FARFAR2 requires a slightly larger subset due to a wider range of predicted \ensuremath{E_{\mathrm{DA}}} values (Figure \ref{fig2}A and B). To ensure a sufficiently large set of structures for subsequent validation while also retaining enough conformers to fully represent the FRET distribution, we chose a structure collection size of 1,000 for each tool. This number is sufficient to model FRET in the unbound (low-FRET) state of the investigated construct but may differ for RNAs with greater structural complexity.

\subsection{Tertiary structure annotations are critical for selecting reliable KLs}

\begin{figure*}[hbt!]%
\centering
\vskip5pt
\includegraphics[width=0.85\textwidth]{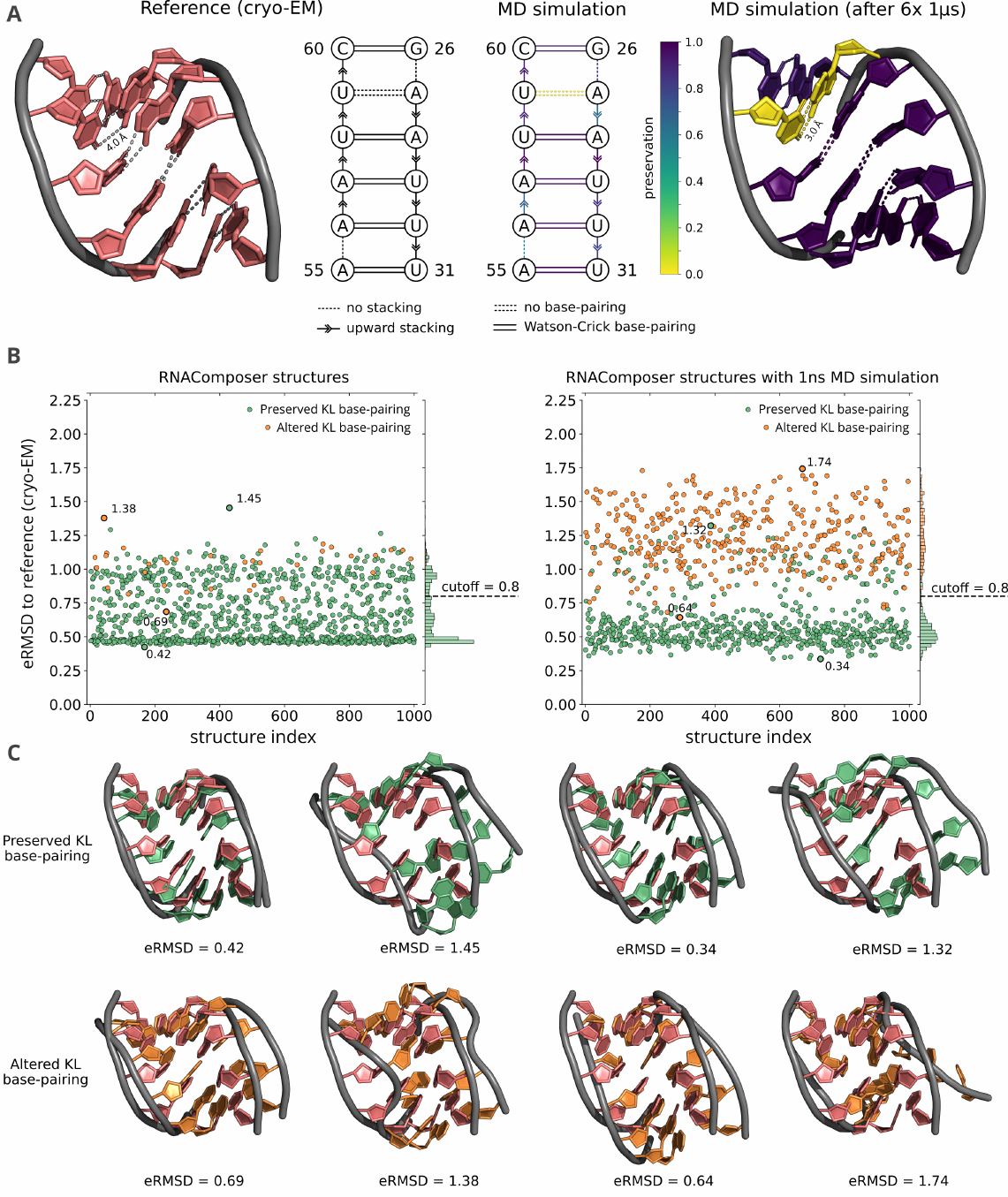}
\caption{Reference KL fold and its comparison with structure collections. \textbf{(A)} The averaged preservation of the KL over the six sets of \SI{1}{\micro\second} MD simulations relative to the reference cryo-EM structure. The results demonstrate conserved WC base-pairings, except for bases A27 and U59, where the conformation observed in the cryo-EM structure is in close spatial proximity but not forming any canonical WC base-pairs. \textbf{(B)} The eRMSD of all structures from the RNAComposer structure collection (left) and RNAComposer structure collection after \SI{1}{\nano\second} MD simulation (right) compared to the reference cryo-EM structure, exclusively focusing on the KL region. \textbf{(C)} The structures classified as preserved KL with the lowest and highest eRMSD, as well as those classified as altered KL, also with the lowest and highest eRMSD with and without MD simulation.}\label{fig3}
\end{figure*}

Since the proper fold of the tertiary contact in our construct depends on a properly foleded KL, we specifically analyzed our structure collections to ensure the accurate formation of the KL in accordance with a cryo-EM reference structure of the ribosomal RNA (PDB-ID: 3JCT). We first assessed the folding of the KL by analyzing the WC base-pairing pattern observed in the reference cryo-EM structure \cite{Leontis2012}. Our analysis revealed that the hydrogen-bonding distance between bases A27 and U59 exceeded the favorable range for stable WC base-pairing and therefore did not fall into any canonical base-pair classification. To investigate the stability and variability of this motif, we used all six \SI{1}{\micro\second} MD simulations of our model construct (Figure \ref{fig3}A). In contrast, these simulations show that all base pairings, including A27–U59, adopt canonical WC geometry, indicating improved pairing consistency in the MD collection. For subsequent classifications of the KL integrity, we established a criterion requiring WC base-pairings between all bases except A27-U59, for which we also allowed spatially proximal representations not classified under a specific WC base-pairing category.

To complement WC base-pairing analysis and refine structural characterization of the KL, we additionally computed the eRMSD for all predicted structures relative to the cryo-EM reference (PDB-ID: 3JCT). The eRMSD quantifies geometric differences between base pairs and stacking interactions, making it a robust metric for assessing RNA motif similarity. Across all prediction tools, a substantial number of structures clustered around an eRMSD of approximately 0.5 (Figure \ref{fig3}B left and Supplementary Figure 4 and 5 left). RNAComposer and FARFAR2 showed smoother transitions to higher eRMSD values, whereas AlphaFold3 exhibited a distinct cutoff. To introduce minor conformational variation and assess structural consistency, each structure was subjected to a short \SI{1}{\nano\second} MD simulation, followed by re-annotation of base pairings and recalculation of the eRMSD (Figure \ref{fig3}B right and Supplementary Figure 4 and 5 right). These simulations introduced local fluctuations in the KL, shifting some conformers above an $\text{eRMSD} = 0.8$ and increasing the number of non WC base-pairs. 

Based on WC base-pairing analysis and eRMSD evaluation, we find that neither metric alone provides a sufficient criterion for verifying a preserved KL formation in our construct based on the reference structure. Using only the threshold of $\text{eRMSD} \leq 0.8$ would retain structures that deviate from the canonical WC pattern and thus do not represent a preserved KL. Conversely, relying solely on WC base-pairing can include structures with significantly altered backbone geometries. However, applying both filters in combination consistently selects folded KL formations adapting WC base-pairing across all three structure prediction tools (Figure \ref{fig3}C and Supplementary Figure 4C and 5C).

\subsection{The experimental smFRET distribution guides the selection of structures from the predicted collections}

\begin{figure*}[hbt!]%
\centering
\includegraphics[width=0.95\textwidth]{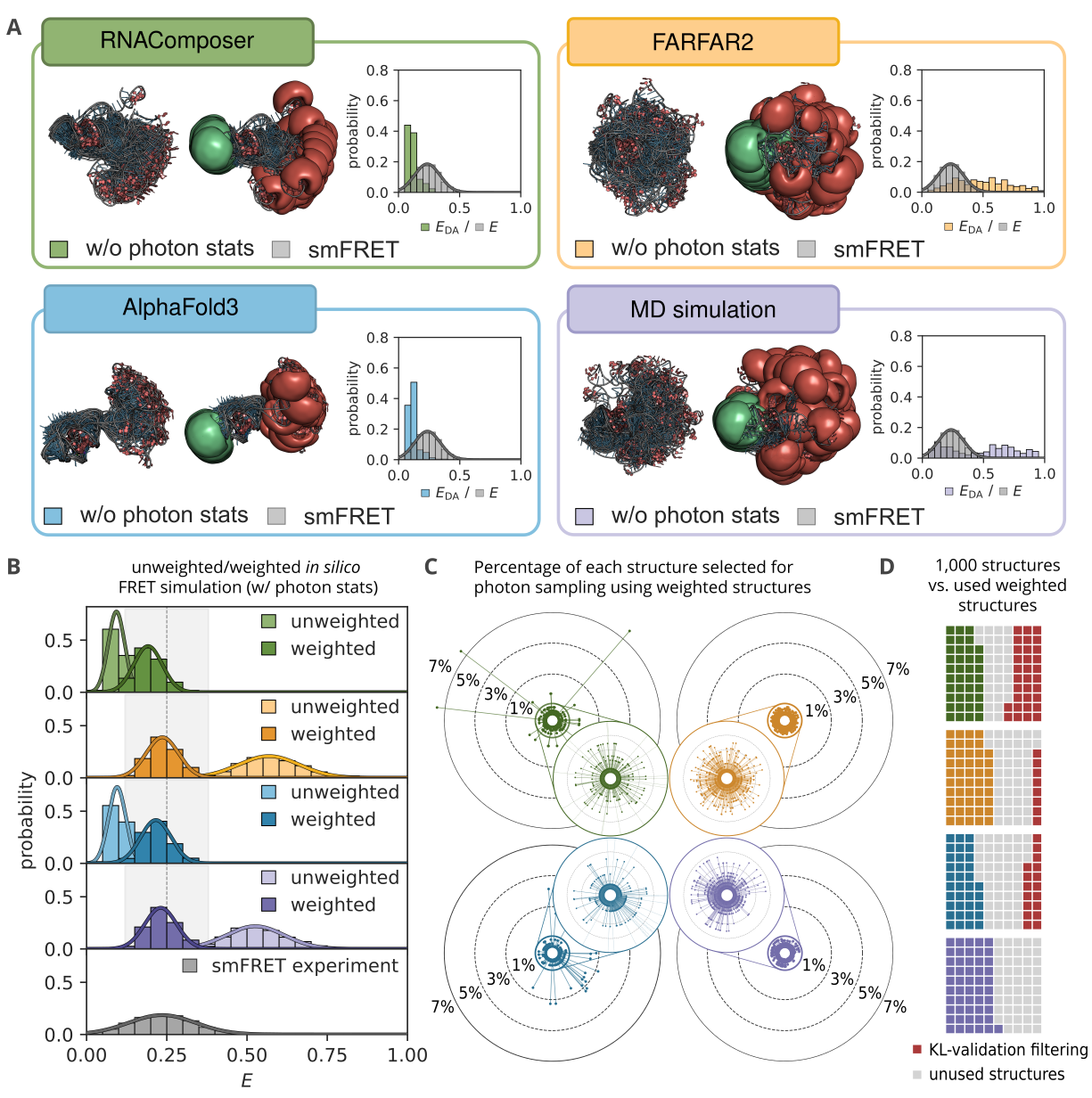}
\caption{Comparison of all 3D structure prediction tools and MD simulations to the smFRET experiment. \textbf{(A)} The FRET distribution of all approaches without photon sampling, as well as the final 3D structure collection after filtering with and without ACVs. \textbf{(B)} Representation of all FRET distributions with photon statistics using FRETraj. "Unweighted" describes a uniform distribution of each structure for the burst calculation, whereas "weighted" denotes the sampling of structures with the same probabilities as in the smFRET experiment. \textbf{(C)} The contribution of all structures for each tool used for the new weighted approach. FARFAR2 has many different structures contributing to the resampled collection, while RNAComposer and AlphaFold3 exhibit few, highly represented structures in the refined collection. \textbf{(D)} The number of lost structures during filtering for validated KLs and, ultimately, how many structures are selected for the weighting approach against the experimental data. KL filtering was not applied to MD simulation structures.}\label{fig4}
\end{figure*}

Building on the filtered structure collections, we can now probe their conformational landscape and guide further refinement using the experimental smFRET distribution. Even without photon simulations, the initial \ensuremath{E_{\mathrm{DA}}} distributions derived from ACVs of all filtered structures already reveal distinct conformational profiles across the different prediction strategies (Figure \ref{fig4}A). FARFAR2 and the combined MD simulations sample a broad range of inter-dye distances, covering not only the experimental smFRET range but the full theoretical FRET range from 0 to 1. This is also evident in the structural visualizations, which display an almost globular distribution of different TL\textsubscript{GAAA} positions. In contrast, RNAComposer and AlphaFold3 predominantly yield conformations in the low-FRET region, suggesting more compact or structurally constrained collections, which are likely due to similarly folded poly(A)-linkers that limit the accessible positions of the TL\textsubscript{GAAA}. Interestingly, these initial distributions already indicate that none of the tools can fully reproduce the experimentally observed broad low-FRET range distribution.

To directly compare our experimental smFRET distribution with the predicted structure collections, we performed photon burst simulations for each structure collection, including all RNA 3D prediction tools and MD trajectories. This approach includes experimental conditions such as shot noise broadening, direct excitation, and gamma correction of the underlying energy transfer efficiency \ensuremath{E_{\mathrm{DA}}} distribution resulting in an \textit{in silico} FRET distribution. Given the assumption of fast dynamics of the RNA model construct in the FRET experiment, bursts were determined by averaging rather than segmenting them into trajectory splits of individual MD runs or subsamples of predicted structures. The resulting FRET distributions are thus uniformly sampled from all structures (Figure \ref{fig4}B - unweighted and Table \ref{table:fret} - w/ photon sampling unweighted). Notably, despite partial overlap, none of the \textit{in silico} FRET distributions generated from RNA 3D prediction tools or MD simulations again fully recapitulate the experimental smFRET distribution.

\begin{table}[h]
\caption{FRET measurements for all approaches}
\label{table:fret}
\begin{tabular}{@{}lccc@{}}
\toprule
              & \multicolumn{1}{l}{w/o photon sampling} & \multicolumn{2}{r}{w/ photon sampling} \\ \midrule
              & \multicolumn{1}{l}{}                    & unweighted         & weighted          \\ \midrule
RNAComposer   & 0.12 $\pm$ 0.05                         & 0.10 $\pm$ 0.03    & 0.20 $\pm$ 0.04   \\
FARFAR2       & 0.50 $\pm$ 0.22                         & 0.58 $\pm$ 0.10    & 0.24 $\pm$ 0.05   \\
AlphaFold3    & 0.12 $\pm$ 0.06                         & 0.10 $\pm$ 0.03    & 0.22 $\pm$ 0.05   \\
MD simulation & 0.48 $\pm$ 0.27                         & 0.53 $\pm$ 0.09    & 0.23 $\pm$ 0.05   \\ \midrule
smFRET        & -                                       & \multicolumn{2}{c}{0.25 $\pm$ 0.13}    \\ \bottomrule
\end{tabular}
\end{table}

Therefore, as the initially predicted \ensuremath{E_{\mathrm{DA}}} distributions without photon sampling appeared promising (Figure \ref{fig4}A), we investigated an alternative sampling approach in which structures are no longer uniformly sampled, but directly
selected based on their probability of appearing in the experimental smFRET distribution. The resulting FRET distributions (Figure \ref{fig4}B - weighted) now closely match the range of the experimental smFRET distribution across all prediction tools and MD simulations (Table \ref{table:fret}, w/ photon sampling - weighted). However, the experimental width of the FRET distribution could not be fully reproduced, despite accounting for shot noise broadening of the experimental burst size distribution and applying gamma correction.

Although our weighted sampling approach successfully reproduced the experimental smFRET distribution, we observed substantial differences in the structural diversity of the selected collections. Particularly for RNAComposer and AlphaFold3, specific FRET ranges are represented by only a few structures. To match the bin probabilities of the experimental distribution, these structures are selected disproportionately often, with some individuals contributing over 7\% to the final weighted RNAComposer collection (Figure \ref{fig4}C). While this enables an accurate match to the experimental FRET profile, it comes at the cost of reduced conformational diversity. The number of structures ultimately contributing to the weighted distribution depends on how many initially fall within the experimental smFRET range and how frequently each FRET bin is populated. FARFAR2 and the MD-based collection show broader structural representation, while AlphaFold3 and RNAComposer rely heavily on a few dominant conformers. The overall filtering process, from the initial structure count through the KL divergence filtering to the final number of structures used in the weighted sampling, is summarized in Figure~\ref{fig4}D (Supplementary Table for exact values).

\section{Discussion}\label{sec3}

FRET-guided integrative modeling has often relied on MD simulations to generate structure collections for comparison with experimental data \cite{Dimura2016, Erichson2021, Steffen2024}. In this study, we have addressed whether RNA 3D prediction tools can serve as a practical alternative, how many structures are actually needed to represent low-FRET collections adequately, and how diverse and reliable the resulting \textit{in silico} FRET predictions and structures are.

We first find that RNAComposer and AlphaFold3 model the unbound state in an extended conformation, leading to a narrow distribution of low energy transfer efficiencies and limited overlap with the experimental smFRET data. In this construct, the TL\textsubscript{GAAA} is connected to the KL via a poly(A)-linker and does not form interactions with either element \cite{Gerhardy2021}. The observed dye distance and the according FRET efficiency is therefore likely influenced by the conformational flexibility of the linker \cite{Hodak2005, Sindbert2011, Barth2022}. Poly(A)-linkers are known to adopt stacked base conformations in low energy states \cite{Downey2006, Willemsen1971}, which may restrict the TL\textsubscript{GAAA} range of motion, consistent with the conformations predicted by RNAComposer and AlphaFold3.

In contrast, FARFAR2 generates a highly diverse structure collection spanning the full FRET range, overlapping the experimental smFRET distribution. This includes poly(A)-linker conformations that allow greater spatial freedom of the TL\textsubscript{GAAA} and therefore higher FRET efficiencies. This indicates that, despite the initial conditions, the poly(A)-linker explores a wide range of conformations on the nanosecond timescale (Supplementary Figures 2 and 3). Similarly, the six \SI{1}{\micro\second} MD trajectories, each sampled from diverse seed conformations, produce a broad FRET distribution. However, since states were randomly selected across all trajectories, the resulting collection is highly dependent on the manually chosen seed structures to ensure representative sampling.

To evaluate how well predicted collections reflect experimentally observed FRET signals under realistic conditions, we simulated photon bursts using FRETraj while accounting for experimental conditions. These corrections ensure that the simulated FRET values reflect what would be observed in a real experiment. FRETraj treats predicted structures, whether from MD simulations or RNA 3D prediction tools like FARFAR2, as if they were rapidly sampled during the photon detection time. This means the method assumes that the RNA explores all conformations on the nanosecond timescale and averages the resulting FRET signals accordingly. As a result, the simulated FRET distributions from both MD and FARFAR2 converge on a similar mean FRET value of approximately 0.5 \cite{Steffen2021, Steffen2024, Hoefling2013}. This convergence indicates, that both structural diversity (static heterogeneity) and fast conformational motion (dynamic averaging) can lead to the same average FRET signal if the timescale of motion is fast compared to the burst detection time window. However, this behavior does not match the experimental FRET distribution, which shows a lower mean and broader width, suggesting that the actual RNA dynamics occur on slower timescales, from microseconds to milliseconds \cite{Schuler2013, Nir2006, Kalinin2010}, which are not captured by the current simulation approach.

We introduced a FRET-guided reweighting strategy, that was implemented to refine the predicted structure collections based on their agreement with the experimental FRET distribution. This procedure yields structure collections, that are not only consistent with the experimental mean (Table \ref{table:fret} w/ photon sampling - weighted) but also reflect the relative occupancy of FRET states. Following reweighting, RNAComposer and AlphaFold3 yield FRET distributions with mean values and widths comparable to those of FARFAR2 and MD, although all remain slightly shifted toward lower efficiencies relative to the experimental distribution. Among these, FARFAR2 best reproduces the experimental mean of 0.25, though the distribution remains narrower than in the experimental data. These findings give rise to two important questions: first, what accounts for the substantial exclusion of structures, particularly from the extensive MD and FARFAR2 collections, during the reweighting process; and second, why does the weighted collection, despite improved agreement in mean FRET efficiency, still fail to recapitulate the full width of the experimental distribution?

The weighting procedure inherently filters out structures with FRET values outside the narrow experimental range, particularly compact conformations with high donor–acceptor proximity. FARFAR2 and MD simulations generate broad collections that include many such structures, which are systematically excluded. The same applies to RNAComposer and AlphaFold3, which contain a large number of low-FRET structures, that despite offering some structural diversity within specific FRET bins, provide limited coverage of the full experimental distribution. It is worth noting that structural properties beyond global distance metrics, such as topological features, may further influence the effective usability of predicted models. In this context, structural entanglements were found to be common across all methods, potentially contributing to the presence of sterically or topologically invalid conformations (Supplementary Tables).

We showed that none of the validated structure collections, including those derived from different MD states, will reproduce the low FRET distribution observed in the smFRET experiment without weighting. To capture the highly dynamic nature of RNA, which is essential for understanding its full conformational landscape \cite{Ganser2019}, thus reconciling the structure heterogeneity across different methodologies and timescales, structure prediction tools must deliver a diverse collection of structures. The latter enables sampling and selecting those structures present in the experimental FRET distribution.

FRET integrative modeling plays a crucial role in elucidating structure dynamics and conformational heterogeneity in biological systems. We advocate that 3D prediction tools, when combined with structural validation, offer a fast alternative to full-atom MD simulations. Furthermore, we demonstrate how FRET can be used to tailor structure collections to experimental observations, allowing for an adaptive adjustment of the initial number of structures depending on the level of detail in the subsequent structure validation. This approach will help shift the perspective from a static reference structure to a dynamic representation of RNA conformational landscapes.
\newline

\noindent\textbf{Acknowledgments}\newline
The authors thank Marika Kaden for her critical feedback regarding the KL divergences and reweighting FRET distributions. Access to smFRET laboratory resources provided by Christian G. Hübner is gratefully acknowledged. The authors MW, FE, and RB gratefully acknowledge financial support by the European Social Fund Plus for Germany (ESF, Grant No. 100649226,  to RB), the German Research Foundation (DFG, Grant No. 498128362 and 537331625 to RB), the Mittweida University of Applied Sciences and Laserinstitut Hochschule Mittweida for further financial support. This project is co-financed from tax revenues based on the budget adopted by the Saxon State Parliament and co-funded by the European Union. MA, TZ and MS acknowledge the support from the Poznan University of Technology and the Institute of Bioorganic Chemistry of the Polish Academy of Sciences (statutory funds), as well as the National Science Centre, Poland (Grant No. 2024/53/B/ST6/02789 to MS). We acknowledge support by the German Research Foundation (DFG) and the Open Access Publication Fund of Mittweida University of Applied Sciences for the OA publication fees. 
\newline
\newline
The funders had no role in study design, data collection and analysis, decision to publish, or manuscript preparation.
\newline

\noindent\textbf{Contributions}\newline
R.B. conceived the research. M.W., F.E. and M.A. generated structure collections of RNA 3D structures. MA and T.Z. analyzed RNAComposer-predicted models for entanglements. M.W. analyzed and compared the structure collections. M.W., F.E., and F.D.S. designed and implemented the software pipeline including \textit{in silico} labeling, MD simulations, FRET predictions incl. mACV dye models, and the software development on Github. R.B. performed and analyzed single-molecule/ensemble fluorescence measurements. M.S. and R.B. supervised the project. M.W. and R.B. wrote the initial draft of the manuscript. All authors contributed to the finalization of the manuscript.
\newline

\noindent\textbf{Data availability}\newline
The source code of FRETraj is available on GitHub \href{https://github.com/RNA-FRETools/fretraj}{https://github.com/RNA-FRETools/fretraj} and Zenodo \href{https://doi.org/10.5281/zenodo.15041891}{https://doi.org/10.5281/zenodo.15041891}. Documentation can be found at \href{https://rna-fretools.github.io/software/}{https://rna-fretools.github.io/software/}. All 10,000 initial structures predicted by RNAComposer, FARFAR2, and AlphaFold3, the downsampled sets of 1,000 structures used for FRET analysis, the six 1 $\mu$s MD trajectories initiated from selected seed structures, and all associated ACV files (.pkl) generated via FRETraj are available at Zenodo: \href{https://doi.org/10.5281/zenodo.15971163}{https://doi.org/10.5281/zenodo.15971163}. 
\newline
\newline
\noindent Additional source material is available from the corresponding author upon request.
\newline

\noindent\textbf{Conflict of interest}\newline
None declared.

\printbibliography

\end{document}


\maketitle

\section*{Supplementary Methods}

\subsection*{smFRET}
To ensure kissing loop formation of the KL-TL$_{\textnormal{GAAA}}$ rRNA model construct, the fluorescently labelled RNA (purchased from IBA) was imaged free in solution in presence of 116 mM KCl, buffered in 50 mM Tris-HCl at pH 7.5 and room temperature \cite{Gerhardy2021}. The single-molecule measurement was performed on a home-built confocal microscope equipped with a $N_\textnormal{A}$ = 1.2 water-immersion objective (Nikon), a Timeharp 260 TSCPC card (Picoquant), a 532 nm cw and a 638 nm pulsed laser source. The latter uses a 10 MHz repetition rate for pulsed overleaved excitation (POE) allowing molecular sorting.\cite{Wahl.2013} Photons were spectrally separated donor 582/64 and acceptor channel 690/70 band pass filters and detected onto two avalanche photodiodes (Perkin Elmer). 
The data was analysed with a sliding time-window with two-color, all photon burst search (APBS) identified bursts with a total count of at least 40 photons ($N_\textnormal{D} + N_\textnormal{A}$). Double labeled molecules were selected by applying an intensity threshold of red photons after acceptor excitation ($N_\textnormal{A,A} > 40$) to remove the donor-only population and a stoichiometry limit $S > 0.2$ to eliminate any acceptor-only species \cite{Kapanidis.2004}. FRET histograms were corrected according to standard protocols\cite{Hellenkamp2018,Lee.2005}. Correction factors are summarized in table \ref{tab:FRETcorrections}.

\subsection*{Structure collection prediction with RNAComposer}

To obtain 10,000 unique \emph{in silico} models of the KL-TL\textsubscript{GAAA} construct \cite{Gerhardy2021}, we used the RNAComposer web server in batch mode (available to logged-in users) at \href{https://rnacomposer.cs.put.poznan.pl/}{https://rnacomposer.cs.put.poznan.pl/} \cite{Sarzynska2023}. We submitted twelve identical batches ($\{batch\_no\}\,i = 1 \ldots 12$), each consisting of $j = 1 \ldots 10$ identical tasks. Each task was defined by a single RNA sequence and ten identical secondary structures $k = 1 \ldots 10$. An example batch is shown below: 

\begin{verbatim}
>KLTL{batch_no}i
#j = 1
UGAAGAAAUUCAAAAAAAAAGCUCGGAAUUUGAGCAAAAAAAAAAAACGGUGGUAAAUUCCAUCG
#k = 1
((((....))))........((((.[[[[[[))))............(((((..]]]]]])))))
...
#k = 10
((((....))))........((((.[[[[[[))))............(((((..]]]]]])))))

...

#j = 10
UGAAGAAAUUCAAAAAAAAAGCUCGGAAUUUGAGCAAAAAAAAAAAACGGUGGUAAAUUCCAUCG
#k = 1
((((....))))........((((.[[[[[[))))............(((((..]]]]]])))))
...
#k = 10
((((....))))........((((.[[[[[[))))............(((((..]]]]]])))))
\end{verbatim}

The input was submitted 1,200 times across all batches (10 inputs per task × 10 tasks per batch × 12 batches). For each such input, RNAComposer generated a family of ten 3D structure collections. The first model in each family was constructed using the highest-scoring structural elements selected deterministically based on RNAComposer’s internal scoring function. The remaining nine models were generated by randomly assembling well-scoring alternative elements (where sequence similarity is still highest, but we allow for, e.g., lower experimental resolution), making the procedure non-deterministic. Consequently, for two identical inputs, the first predicted model would be the same, while the remaining nine would, with high probability, differ. As a result of processing 1,200 identical inputs, we obtained 12,000 RNA 3D models, of which 1,200 were identical (the first models from each family) and 10,800 were structurally diverse. The corresponding ZIP files containing all 12,000 models were downloaded from the RNAComposer web server workspace and stored in the '\emph{models}' folder. Filtering out identical structures was performed at a later stage of the processing pipeline. From the downloaded archives, we extracted only the PDB files and then removed the ZIP files using the following command:

\begin{verbatim}
find . -name "*.zip" -exec unzip -d . -j {} "*/*.pdb" \;; rm *.zip
\end{verbatim}

\noindent We then split multi-model PDB files into individual model files using the following AWK script:

\begin{verbatim}
BEGIN { filename = ARGV[1]; gsub(/" "/, "-",filename);
split(filename, elems, ".");}
$1 == "MODEL" {file = elems[1] "#" $2 ".pdb";content="";}
$1 == "ATOM" || $1 == "HETATM" || $1 == "TER" {content = content $0 "\n";}
$1 == "ENDMDL"  {printf("%s",content) > file;}
\end{verbatim}

\noindent The script was saved as \texttt{split} and applied within the '\emph{models}' folder using the command:

\begin{verbatim}
for i in *.pdb; do awk -f split "$i"; done; 
find . -type f ! -name "*#*.pdb" -exec rm {} \;
\end{verbatim}

\noindent To remove duplicate 3D structures from the prediction set, we used the \texttt{fdupes} tool. First, we created a copy of the full set of models in a new folder:

\begin{verbatim}
cp models models-without-duplicates
\end{verbatim}

\noindent Then, we applied \texttt{fdupes} to identify and remove redundant structures:

\begin{verbatim}
fdupes -rdN models-without-duplicates > ./fdupes.log
\end{verbatim}

\noindent This procedure reduced the dataset to 10,801 unique 3D models, which were stored in the '\emph{models-without-duplicates}' folder. From this set, we randomly selected 10,000 structures for further analysis and copied them to the final folder named '\emph{rnacomposer-kl-tlgaaa}' using the following commands:

\begin{verbatim}
mkdir rnacomposer-kl-tlgaaa;
find models-without-duplicates -mindepth 1 -maxdepth 1 -name '*.pdb' -print0 |
shuf -n 10000 -z | xargs -r0 cp -t rnacomposer-kl-tlgaaa;
\end{verbatim}

\noindent These 10,000 unique RNA 3D models predicted by RNAComposer were used in the analyses described in this study.

\subsection*{Structure collection prediction with FARFAR2}

For 3D RNA structure prediction of the KL-TL\textsubscript{GAAA} model construct, we used FARFAR2 \cite{Watkins2020} from the Rosetta Commons toolbox (\href{https://rosettacommons.org/}{https://rosettacommons.org/}), incorporating sequence and secondary structure constraints to generate 10,000 individual structures. The structure generation process was adapted from the Supplementary Information of Steffen et al. \cite{Steffen2024}. The following command was executed using \texttt{rna\_denovo}:

\begin{verbatim}
rna_denovo.linuxgccrelease -nstruct 10000 -fasta kltlgaaa_sequence.fasta 
-secstruct_file kltlgaaa_secstruct.txt -silent kltlgaaa.out 
-minimize_rna true -cycles 20000
\end{verbatim}
The contents of \texttt{kltlgaaa\_sequence.fasta} and \texttt{kltlgaaa\_secstruct.txt} files were as follows:
\begin{verbatim}
> KL-TLGAAA construct
ugaagaaauucaaaaaaaaagcucggaauuugagcaaaaaaaaaaaacggugguaaauuccaucg
\end{verbatim}
\begin{verbatim}
((((....))))........((((.[[[[[[))))............(((((..]]]]]])))))
\end{verbatim}
The 10,000 structures were subsequently extracted using \texttt{rna\_denovo} and merged into a single PDB file with the following commands:
\begin{verbatim}
grep "^SCORE:" kltlgaaa.out | grep -v description | awk '{print $NF ": " $2}' | 
tee all_models.txt

extract_pdbs.linuxgccrelease -in:file:silent kltlgaaa.out -tags
`cat all_models.txt | cut -f1 -d':'`

for pdb in `cat all_models.txt | cut -f1 -d':'`; do
  echo MODEL $i >> kltlgaaa.pdb
  echo TITLE "$pdb" >> kltlgaaa.pdb
  cat "$pdb".pdb >> kltlgaaa.pdb
  echo -e "ENDMDL\n\n" >> kltlgaaa.pdb
  i=$(($i+1))
done
\end{verbatim}

\subsection*{Structure collection prediction with AlphaFold3}

For structure prediction using AlphaFold3 \cite{Abramson2024} (\href{https://github.com/google-deepmind/alphafold3}{https://github.com/google-deepmind/alphafold3}), 10,000 structures were computed locally using Docker, utilizing the model weights distributed by AlphaFold for non-commercial use (see installation instructions: \href{https://github.com/google-deepmind/alphafold3/blob/main/docs/installation.md}{https://github.com/google-deepmind/alphafold3/blob/main/docs/installation.md}). The following input JSON file was used to generate 10,000 predicted structures with AlphaFold3, where \texttt{modelSeeds} was defined as a list ranging from 1 to 10,000.

\begin{lstlisting}[language=json]
{
  "name": "KLTL_GAAA",
  "modelSeeds": [1, ..., 10000],
  "sequences": [
    {
      "rna": {
        "id": "A",
        "sequence": "UGAAGAAAUUCAAAAAAAAAGCUCGGAAUUUGAGCAAAAAAAAAAAACGGUGGUAAAUUCCAUCG"
      }
    }
  ],
  "dialect": "alphafold3",
  "version": 2
}
\end{lstlisting}

\subsection*{MD simulations}

Molecular dynamics (MD) simulations were performed using GROMACS 2024.2 \cite{Abraham2015} with the AMBER force field \cite{Cornell1995}, incorporating parmbsc0 \cite{Perez2007} and $\chi$OL3 \cite{Banas2010, Zgarbova2011} corrections for RNA. The RNA was solvated in a dodecahedral water box with TIP3P or TIP4P \cite{Abascal2005}, charge-neutralized with KCl, and equilibrated at 298 K and 1 bar. The individual helices of the model construct (H22, H68, H88) were extracted from the reference structure (PDB: 3JCT \cite{Wu2016}) and connected \textit{in silico} via poly-A linkers using PyMOL (https://www.pymol.org/). 

In total, six MD simulations were conducted using TIP4P water, each lasting 1~$\mu$s. All MD simulations were used for analyzing the kissing loop of the model construct using BARNABA, specifically assessing the preservation of WC base-pairings. The seed structures for the six MD simulations were derived from the cryo-EM reference structure (PDB ID: 3JCT), truncated to the tertiary contact core region comprising helices H22, H68, and H88. Nucleotides within the KL region that deviated from the target sequence were mutated using PyMOL \cite{PyMOL}. 

A poly-A linker connecting H22 and H88 was constructed using PyMOL builder and manually integrated into the model so that it extended behind the KL receptor site. The tetraloop from the reference structure was manually placed at six distinct positions, resulting in six individual structures with identical KL domains. The poly-A linker between H88 and H68 was also generated using the PyMOL Builder and subsequently adjusted to connect each tetraloop configuration to its corresponding KL domain.

\clearpage

\subsection*{FRETraj Dye parameters}

The following FRETraj \cite{Steffen2021} dye parameters were used to compute the mACVs for all structures, serving as the basis for calculating $E_{DA}$ values and generating the corresponding FRET efficiency histograms. Atom IDs had to be adjusted for each 3D prediction tool due to the different numbering of atoms in the PDB files.

\begin{lstlisting}[language=json]
{"Position":
                  {"Cy3-65-O3'":
                       {"attach_id": <ATOM-ID in PDB>,
                        "mol_selection": "all",
                        "linker_length": 20,
                        "linker_width": 3.5,
                        "dye_radius1": 8,
                        "dye_radius2": 3,
                        "dye_radius3": 1.5,
                        "cv_fraction": 0.25,
                        "cv_thickness": 3,
                        "use_LabelLib": false,
                        "grid_spacing": 1.0,
                        "simulation_type": "AV3",
                        "state": 1,
                        "frame_mdtraj": 0,
                        "contour_level_AV": 0,
                        "contour_level_CV": 0.7,
                        "b_factor": 100,
                        "gaussian_resolution": 2,
                        "grid_buffer": 2.0,
                        "transparent_AV": true
                        },
                   "Cy5-10-C5":
                       {"attach_id": <ATOM-ID in PDB>,
                        "mol_selection": "all",
                        "linker_length": 20,
                        "linker_width": 3.5,
                        "dye_radius1": 9.5,
                        "dye_radius2": 3,
                        "dye_radius3": 1.5,
                        "cv_fraction": 0.25,
                        "cv_thickness": 3,
                        "use_LabelLib": false,
                        "grid_spacing": 1.0,
                        "simulation_type": "AV3",
                        "state": 1,
                        "frame_mdtraj": 0,
                        "contour_level_AV": 0,
                        "contour_level_CV": 0.7,
                        "b_factor": 100,
                        "gaussian_resolution": 2,
                        "grid_buffer": 2.0,
                        "transparent_AV": false}
                   },
              "Distance": {"Cy3-Cy5":
                               {"R0": 54,
                                "n_dist": 1000000}
                           }
              }

\end{lstlisting}

\clearpage

\subsection*{FRETraj photon sampling parameters}

The following parameters were used for photon sampling with FRETraj across all structure collections to generate both unweighted and weighted FRET distributions, using the corresponding rkappa files.

\begin{lstlisting}[language=json]
{
    "dyes": {
        "tauD": 1.4,
        "tauA": 1.12,
        "QD": 0.2,
        "QA": 0.35,
        "etaA": 1,
        "etaD": 0.37,
        "dipole_angle_abs_em": 0
    },
    "sampling": {
        "nbursts": 20000,
        "skipframesatstart": 0,
        "skipframesatend": 0,
        "multiprocessing": true
    },
    "fret": {
        "R0": 54,
        "kappasquare": 0.6666,
        "gamma": true,
        "quenching_radius": 1
    },
    "species": {
        "name": ["all"],
        "unix_pattern_rkappa": ["<unweighted> or <weighted> r_kappa file"],
        "unix_pattern_don_coords": [],
        "unix_pattern_acc_coords": [],
        "probability": [1],
        "n_trajectory_splits": null
    },
    "bursts": {
        "lower_limit": null,
        "upper_limit": null,
        "lambda": null,
        "QY_correction": false,
        "averaging": "all",
        "burst_size_file": "<experimental burst sizes>"
    }
}
    
\end{lstlisting}

\clearpage

\section*{Supplementary Figures}

\begin{figure}[h!]
    \centering
    \includegraphics[width=1\linewidth]{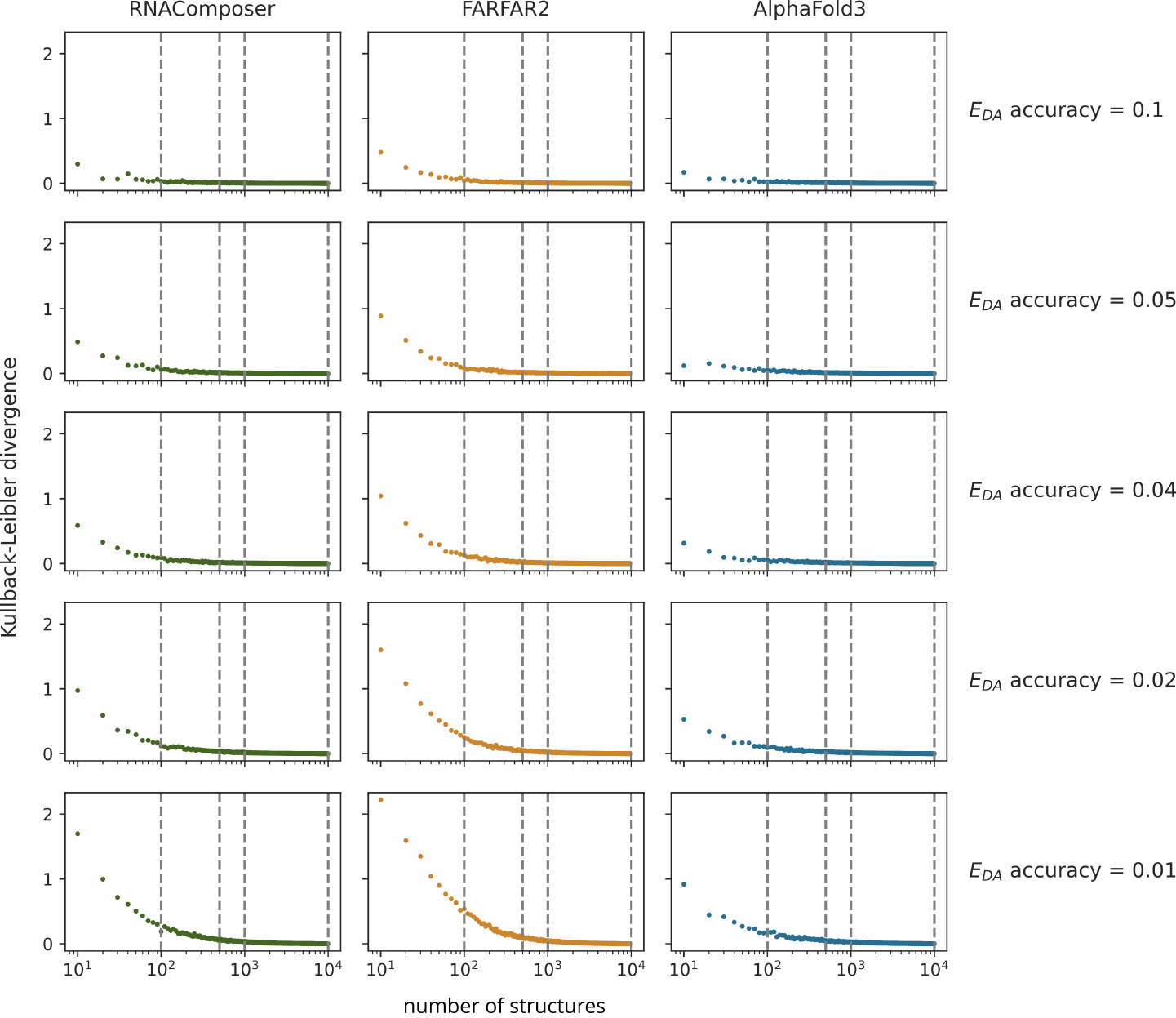}
    \caption{Dependence of Kullback-Leibler divergence (KLD) on the number of structures in the structure collections generated by RNAComposer, FARFAR2, and AlphaFold3. The $E_{DA}$ accuracy reflects the quality of FRET-based predictions and is affected by the bin size used for computing KLD. At an $E_{DA}$ accuracy of 0.1, as few as 100 structures suffice to represent a full structure collection of 10,000 structures across all tools, whereas an accuracy of 0.01 requires around 1,000 structures.}
    \label{fig:SI1}
\end{figure}

\clearpage

\begin{figure}[h]
    \centering
    \includegraphics[width=1\linewidth]{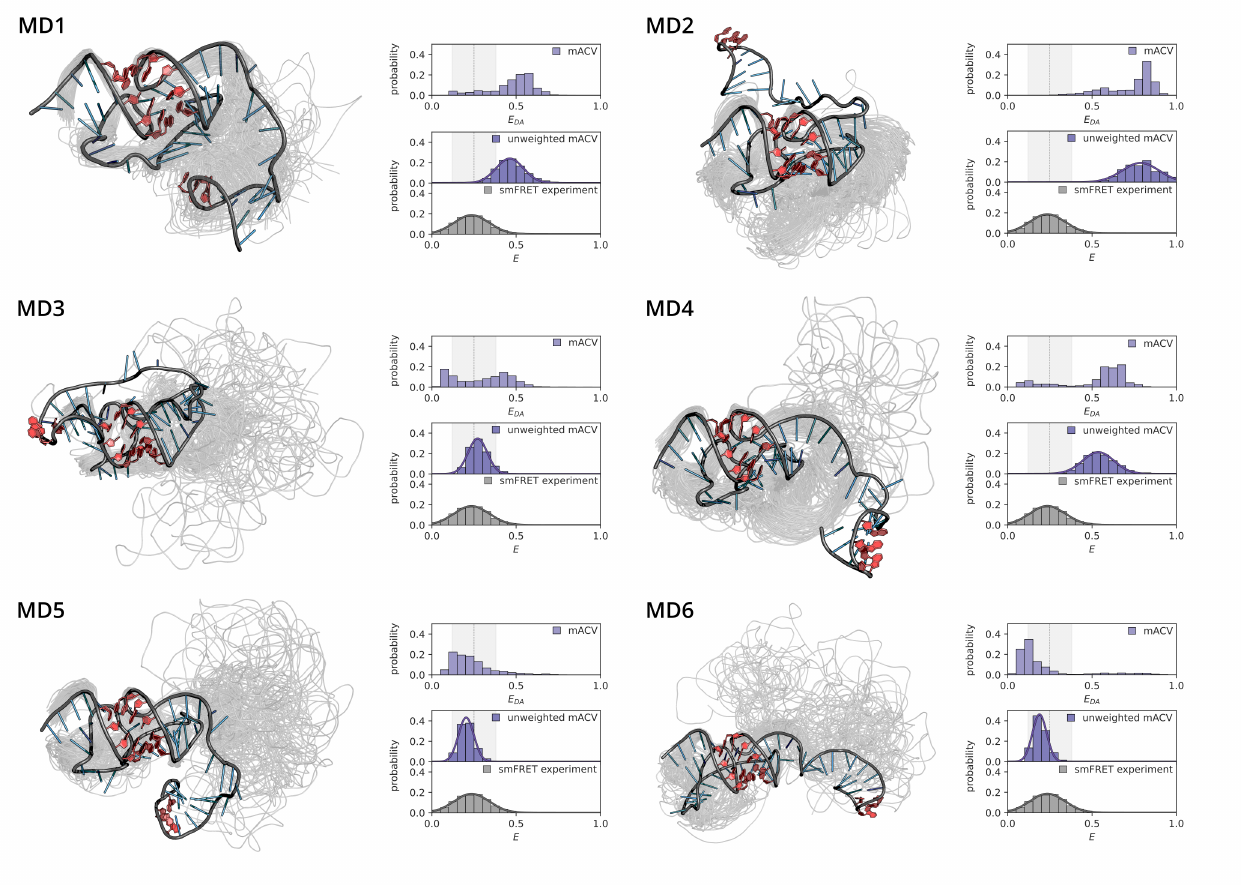}
    \caption{Seed structures used as starting points for the MD simulations. For each simulation, the input structure and ribbon conformations sampled every \unit[10]{ns} are shown. Additionally, the corresponding $E_{DA}$ distributions and unweighted FRET distributions obtained after photon sampling are plotted and compared with the experimental smFRET histogram.}
    \label{fig:SI2}
\end{figure}

\clearpage

\begin{figure}[h]
    \centering
    \includegraphics[width=1\linewidth]{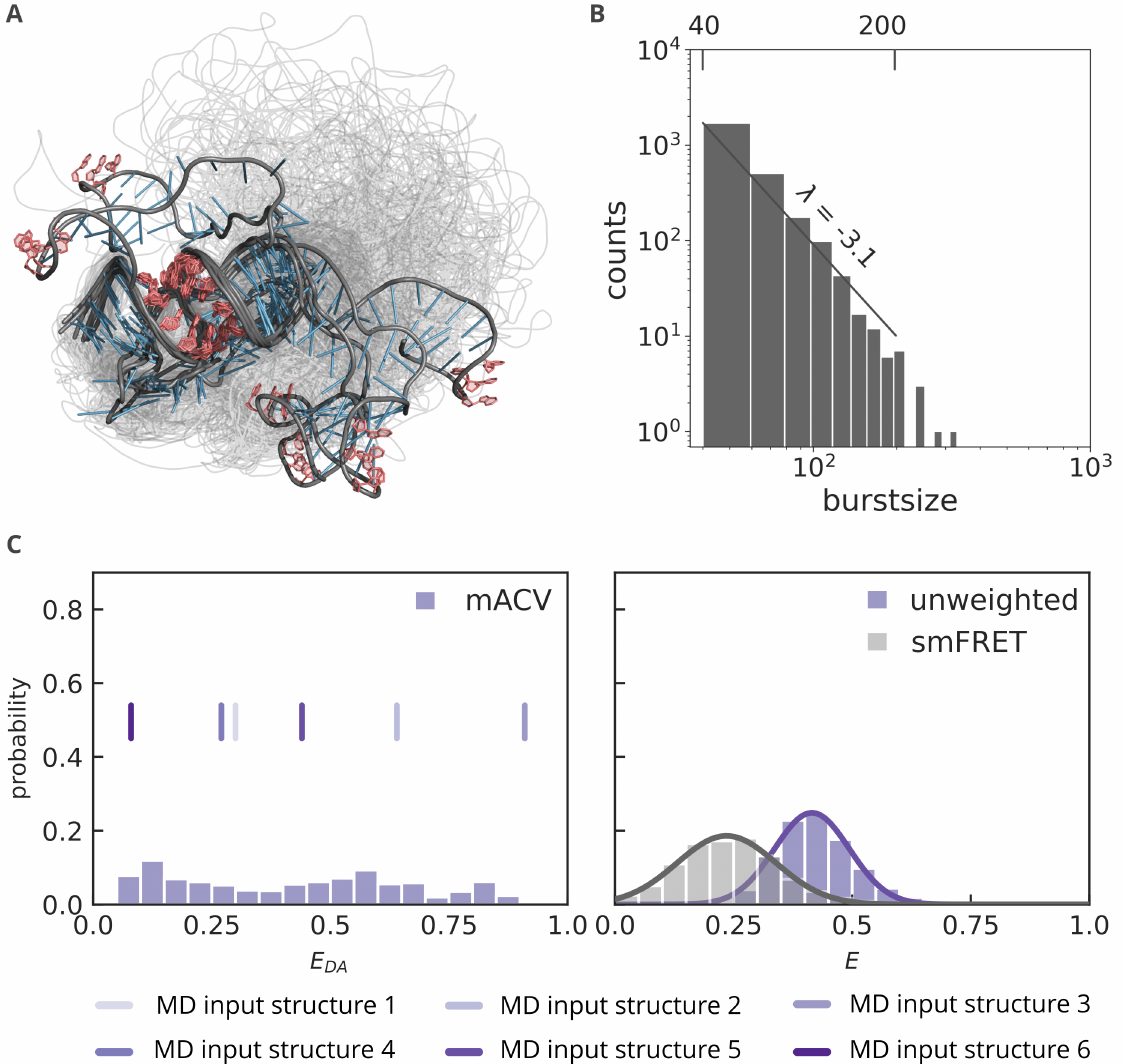}
    \caption{Six seed structures treated as one structure collection, with corresponding simulated FRET distributions. \textbf{(A)} shows the seed structures as well as the states, which were randomly chosen from all six simulations. \textbf{(B)} displays the experimental burst size distribution used for the calculation of the \textit{in silico} burst simulation with FRETraj. \textbf{(C)} shows the $E_{DA}$ histogram for the combined MD simulation states as well as the six input $E_{DA}$ values of the seed structures on the left side. The right side shows the $E$ histogram (with photon sampling) plotted against the experimental smFRET histgoram.} 
    \label{fig:SI2}
\end{figure}

\clearpage

\begin{figure}[h]
    \centering
    \includegraphics[width=1\linewidth]{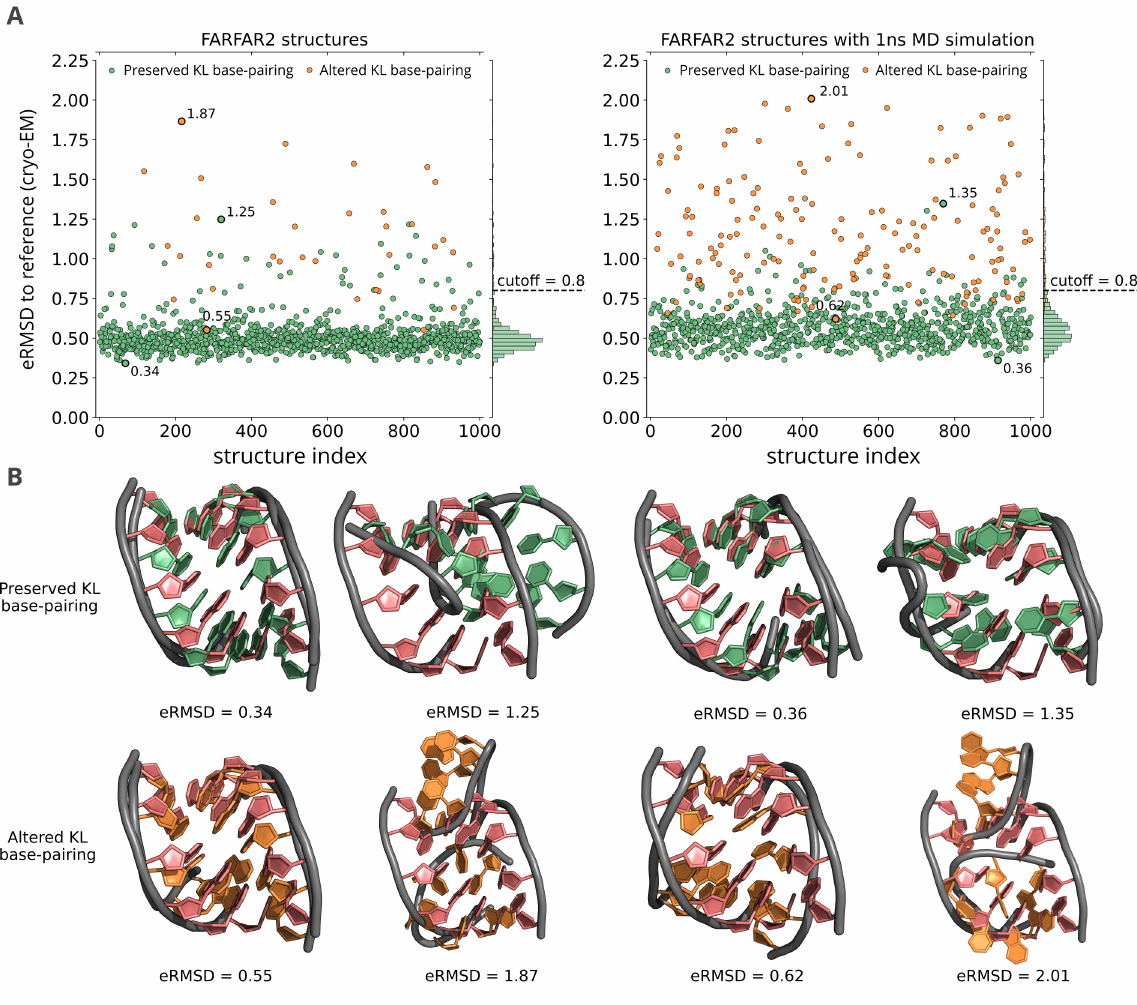}
    \caption{Correct kissing loop formation and eRMSD values for the FARFAR2 structure collection before and after 1ns MD simulation. \textbf{(A)} eRMSD of all structures from the FARFAR2 collection before (left) and after (right)  \unit[1]{ns} MD simulation, relative to the reference cryo-EM structure, focusing on the kissing loop region. \textbf{(B)} Representative structures classified as correctly forming the kissing loop (with the lowest and highest eRMSD), and incorrectly forming the kissing loop (also with the lowest and highest eRMSD), both before and after MD simulation.}
    \label{fig:SI3}
\end{figure}

\clearpage

\begin{figure}[h]
    \centering
    \includegraphics[width=1\linewidth]{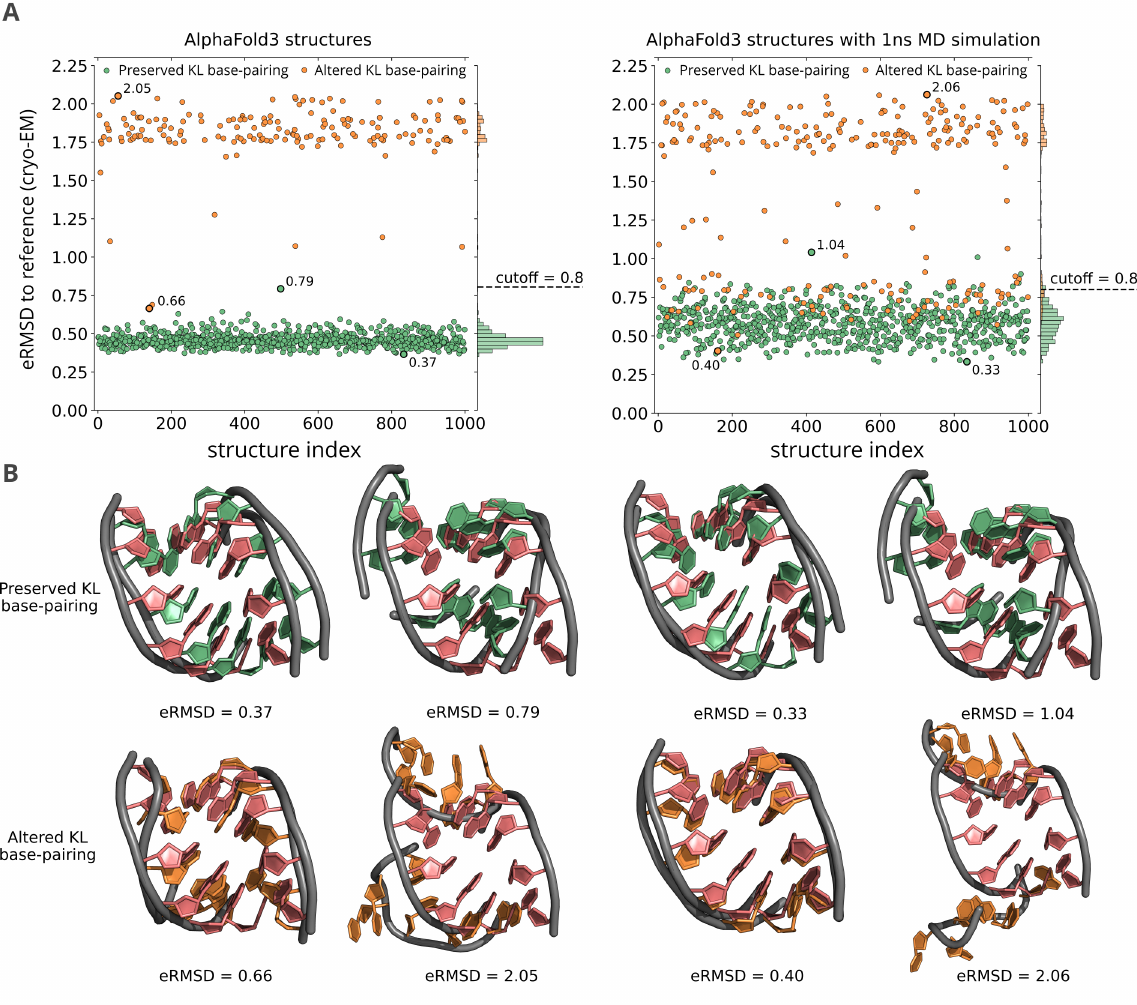}
    \caption{Correct kissing loop formation and eRMSD values for the AlphaFold3 structure collection before and after 1ns MD simulation. \textbf{(A)} eRMSD of all structures from the AlphaFold3 collection before (left) and after (right) \unit[1]{ns} MD simulation, relative to the reference cryo-EM structure, focusing on the kissing loop region. \textbf{(B)} Representative structures classified as correctly forming the kissing loop (with the lowest and highest eRMSD), and incorrectly forming the kissing loop (also with the lowest and highest eRMSD), both before and after MD simulation.}
    \label{fig:SI4}
\end{figure}

\clearpage

\begin{figure}[h]
    \centering
    \includegraphics[width=1\linewidth]{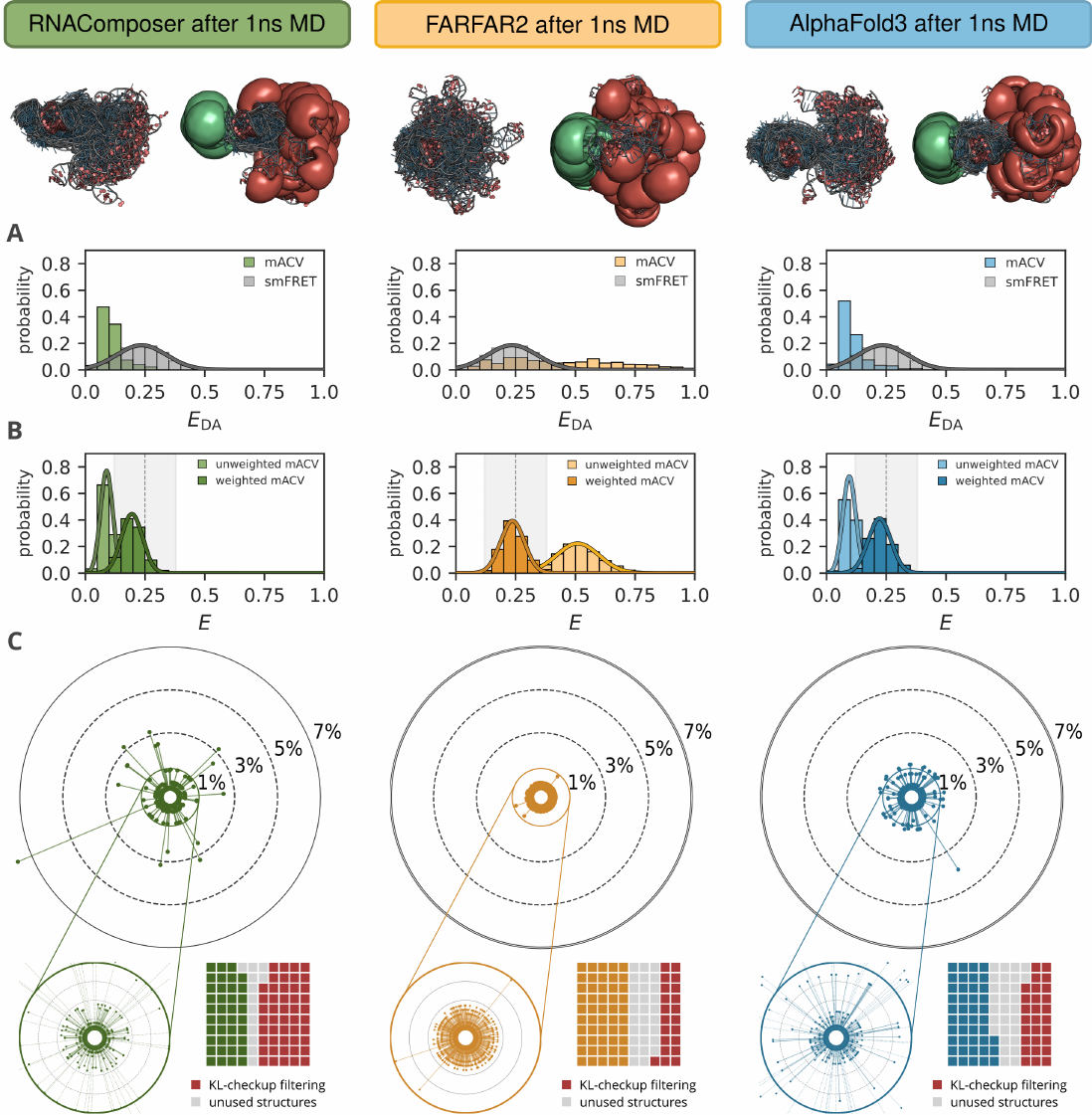}
    \caption{FRET distributions and structural composition of model collections after \unit[1]{ns} MD simulation. \textbf{(A)} FRET distributions obtained from all methods without photon sampling. \textbf{(B)} FRET distributions calculated with photon statistics using FRETraj \cite{Steffen2021}. "Unweighted" refers to uniform structure contributions during burst calculation; "weighted" reflects sampling based on experimental smFRET probabilities. \textbf{(C)} Structure contributions of each tool in the weighted approach. More uniform structure proportions indicate higher structural diversity and a greater number of required models. The waffle chart illustrates the reduction in structure numbers after filtering for validated kissing loops and the final number of structures used in the weighting analysis.}
    \label{fig:SI5}
\end{figure}

\clearpage

\section*{Supplementary Tables}

\begin{table}[htbp]
\centering
\caption{FRET correction factors for spectral crosstalk including bleed-through (bt) and direct excitation (dE), as well as the $\gamma$-correction factor with detection efficiency $\eta$ and quantum yield $Q$.}
\begin{tabular}{llcc}
\hline
\textbf{Category} & \textbf{Correction factor} & \textbf{Donor} & \textbf{Acceptor} \\
\hline
Bleed-through     & $bt$                      & 9.1\%          & 0.3\%             \\
Direct excitation & $dE$                      & 0.2\%          & 4.2\%             \\
Detection efficiency & $\eta$                & \multicolumn{2}{c}{$\eta_A/\eta_D = 2.72$} \\
Quantum yield     & $Q$                       & 0.20           & 0.35              \\
\hline
\end{tabular}
\label{tab:FRETcorrections}
\end{table}

\begin{table}[h]
\centering
\caption{Overview of structure loss due to kissing loop filtering and entanglements found in the remaining structures}
\begin{tabular}{@{}lccc@{}}
\toprule
                                                                     & Initial collection & After KL check & w/o entanglements \\ \midrule
RNAComposer                                                          & 1,000               & 675            & 338               \\ \midrule
\begin{tabular}[c]{@{}l@{}}RNAComposer\\ (after 1ns MD)\end{tabular} & 1,000               & 527            & 269               \\ \midrule
FARFAR2                                                              & 1,000               & 935            & 139               \\ \midrule
\begin{tabular}[c]{@{}l@{}}FARFAR2\\ (after 1ns MD)\end{tabular}     & 1,000               & 803            & 96               \\ \midrule
AlphaFold3                                                           & 1,000               & 829            & 17               \\ \midrule
\begin{tabular}[c]{@{}l@{}}AlphaFold3\\ (after 1ns MD)\end{tabular}  & 1,000               & 716            & 52               \\ \bottomrule
\end{tabular}
\label{tab:structures}
\end{table}

\noindent Table \ref{tab:structures} highlights the importance of filtering implausible artificially generated 3D structures across the three tools. Although RNAComposer loses a large fraction of its models due to eRMSD and Watson–Crick base-pairing filters, only about half of the remaining structures exhibit entanglements. In contrast, FARFAR2 and AlphaFold3 retain most structures after our initial filtering, but nearly all are removed when additionally filtering for entanglements detected by RNAspider \cite{Luwanski2022}.

\printbibliography